\documentclass[a4paper,fleqn,usenatbib]{mnras}
\usepackage[T1]{fontenc}
\usepackage{ae,aecompl}
\usepackage{graphicx}
\usepackage{rotating} 
\usepackage{epsfig} 
\usepackage{amsmath}
\usepackage{amssymb}
\usepackage{euscript}
\usepackage{pdflscape}
\usepackage{url}

\title[The envelope of V CVn ]{The envelope of the semiregular variable V CVn}

\author[B. Safonov et al.]{
Boris Safonov,$^{1}$\thanks{E-mail: safonov@sai.msu.ru (BS)}
Sergei Lamzin,$^{1}$
Alexandr Dodin,$^{1}$
Alexey Rastorguev$^{1}$
\\
$^{1}$Sternberg Astronomical Institute of Lomonosov Moscow State University, Universitetskii pr-t, 13, Moscow 119992, Russian Federation \\
}

\date{Accepted XXX. Received YYY; in original form ZZZ}

\pubyear{2019}

\begin{document}
\label{firstpage}
\pagerange{\pageref{firstpage}--\pageref{lastpage}}
\maketitle

\begin{abstract}
V~CVn is a semiregular variable star with a $V$--band amplitude of $\approx2$ mag. This star has an unusually high amplitude of polarimetric variability: up to 6 per cent. It also exhibits a prominent inverse correlation between the flux and the fraction of polarization and a substantial constancy of the angle of polarization. To figure out the nature of these features, we observed the object using the Differential Speckle Polarimetry at three bands centered on 550, 625 and 880~nm using the 2.5-m telescope of Sternberg Astronomical Institute. The observations were conducted on 20 dates distributed over three cycles of pulsation. We detected an asymmetric reflection nebula consisting of three regions and surrounding the star at the typical distance of 35~mas. The components of the nebula change their brightness with the same period as the star, but with significant and different phase shifts. We discuss several hypotheses that could explain this behavior.
\end{abstract}




\section{Introduction}

The radiation of red long--period variables (LPV) is often polarized due to scattering on the dust, which is being formed in their relatively cool atmospheres. The fraction and angle of polarization of LPV fluctuate randomly due continuous chaotic changes in the envelopes of these object. Usually the fraction of polarization changes from 0 to 2 per cent, the orientation of the polarization plane has no preferred direction \citep{ClarkeBook}. A semiregular variable star V~CVn, which has a period of $194$~days \citep{Samus2017} and $V$--band amplitude of $\approx2$~mag, stands out against this background. The fraction of polarization of this star changes from 1 to 6 per cent while the angle of polarization is quite stable: from $99^{\circ}$ to $122^{\circ}$ \citep{Serkowski2001}. In addition, V~CVn exhibits most prominent inverse correlation between flux and fraction of polarization among other long--period variables.

\citet{Neilson2014} discussed the polarization variability of V~CVn in detail. They considered qualitatively several hypotheses which can potentially describe unique behavior of the star. They concluded that the model of dusty disc and the model of bow shock are the most probable. In the first case, the intrinsic polarization is generated by the scattering from the dusty thick disc or torus. The observer is close to the plane of equator of this structure. The second hypothesis states that a bow shock is formed at the boundary between the stellar wind of V~CVn and the interstellar medium, similar to one found around $o$~Ceti \citep{Martin2007}. Dust from the wind will be accumulated at this boundary. This dusty structure will also scatter and polarize stellar radiation.


In both cases an asymmetry of scattering envelope emerges. It can potentially produce non--zero intrinsic polarization of the object and the constancy of polarization angle. \citet{Neilson2014} showed how the inverse correlation between the total flux and polarization of the object can be qualitatively explained by the interaction between the pulsation--driven density waves and the bow shock or dusty disk. 




\begin{figure*}
	\includegraphics[width=2.1\columnwidth]{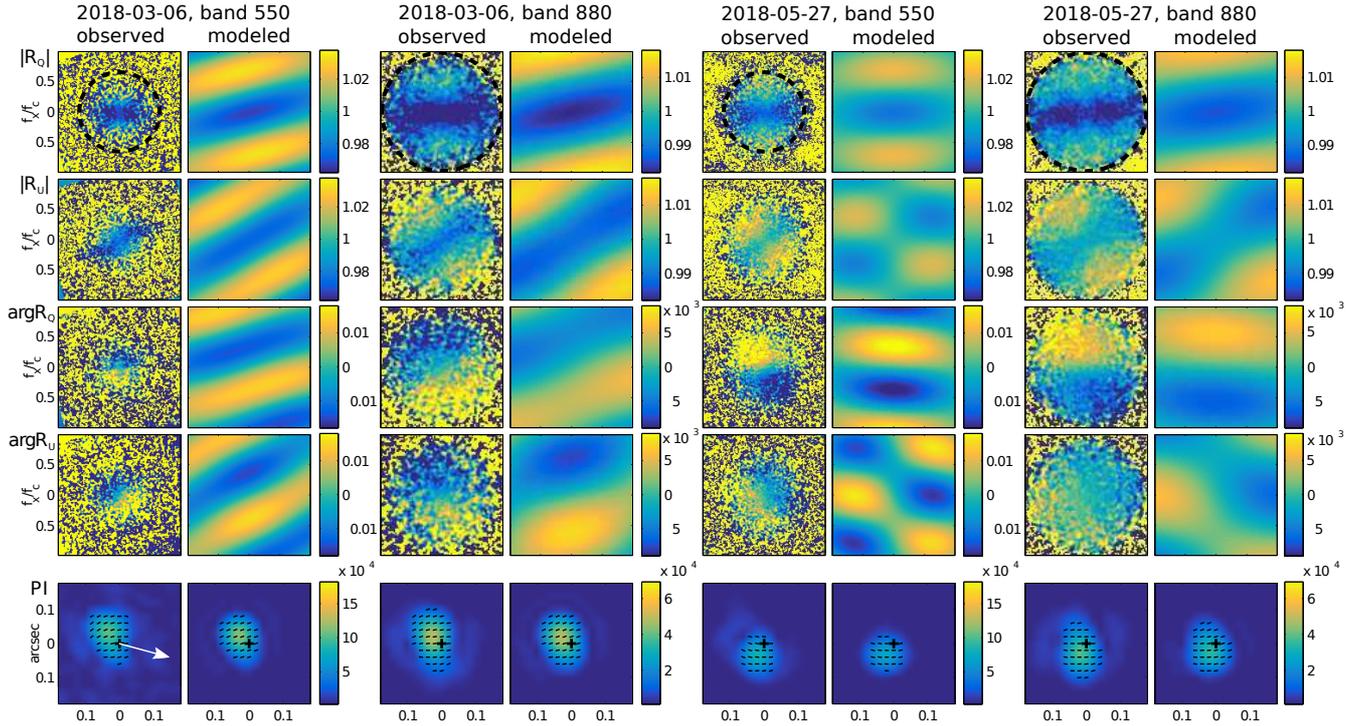}
    \caption{The results of observations of V~CVn (odd columns) and their modeling (even columns): the data for two dates and at two filters (see the titles of panels in the upper row). The visibility ratios in the orthogonal polarizations (\ref{eq:RvalDef}) are presented in four upper rows : $|\mathcal{R}_{Q}|$, $|\mathcal{R}_{U}|$, $\mathrm{arg}\mathcal{R}_{Q}$, $\mathrm{arg}\mathcal{R}_{U}$. The spatial frequency divided by the cut--off frequency $D/\lambda$ is along the axes. In the lower row the polarized flux in pixel divided by the total flux of the object is presented (the angular size of pixel is 20.6 mas/pix). The short lines depict the polarization orientation, they have arbitrary length. The angular coordinates are along the axes. Dashed circles in the upper row depict the region in the Fourier space where the comparison between measurements and models took place (see Section \ref{sec:model}). The arrow in the lower left panel indicates the proper motion of the star (see Section \ref{sec:interp}). In all panels, North is up and East is left.}
    \label{fig:VCVn_RVal}
\end{figure*}

The existing array of polarization measurements of V~CVn covers dozens of pulsation cycles, what allows to state that the peculiar behavior of the object is being reproduced. However in these measurements the polarization of the whole object was integrated hampering further interpretation. The localization of polarized flux source or sources may be the key to understanding of the object.

Here we report on the observations of V~CVn using a high angular resolution polarimetry technique. We detected a circumstellar environment around the star, which behaviour may explain polarimetric variability of the object. The paper is organized as follows. We briefly describe a method and observations in Section \ref{sec:observations}. In Section \ref{sec:model} we construct a simple geometric model of the observations. The discussion of possible interpretations and conclusions are provided in Sections \ref{sec:interp} and \ref{sec:conclusions}, respectively.

\section{Observations}
\label{sec:observations}

V~CVn was observed using the SPeckle Polarimeter (SPP) of the 2.5-m telescope of the Caucasian Observatory of the Sternberg Astronomical Institute of Lomonosov Moscow State University. The SPP is a combination of a two--beam polarimeter and a speckle interferometer \citep{Safonov2017}. The instrument is aimed at the study of polarization of astrophysical objects at diffraction limited angular resolution, i.e. 50~mas at wavelength of 500~nm. The angular scale of SPP camera is 20.6~mas pix$^{-1}$.

The observations were conducted on 18 dates distributed from 2017 December 2 to 2019 January 20 in the fast polarimetry regime using three medium--band filters centered on 550, 625, and 880~nm. In addition, we observed the object in filters $V$ and $I_\mathrm{c}$ on two dates in the spring of 2017.

\begin{figure*}
\begin{center}
	\includegraphics[width=2.1\columnwidth]{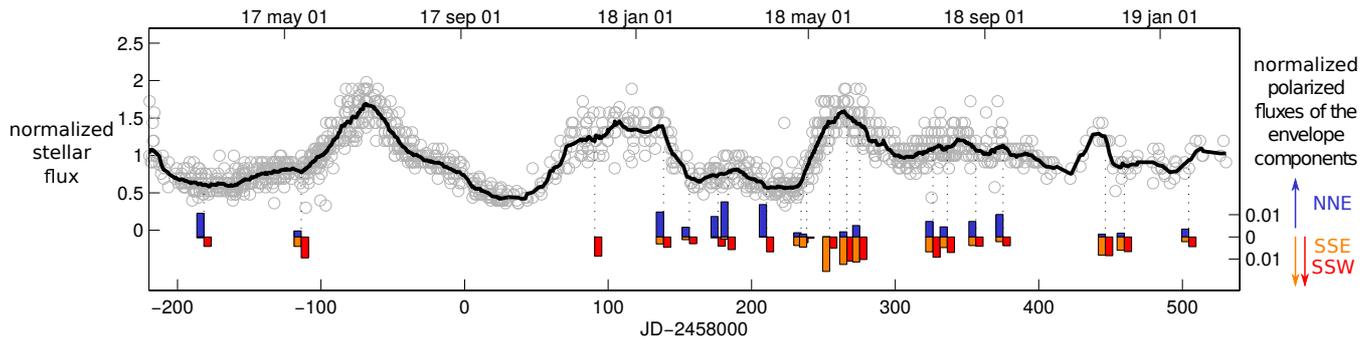}
\end{center}
    \caption{
    Individual AAVSO estimations of visual magnitude converted to flux and normalized by its average are indicated by grey circles. The thick black line indicates a running average of the AAVSO fluxes (the right $OY$ axis). Coloured bars in the lower part of figure stands for the polarized fluxes of the envelope components, normalized by the average total flux of the object. Each observation is displayed by a group of three bars: blue for NNE arc, orange for SSE arc, red for SSW arc. The length of bar indicates polarized flux of the corresponding arc of the envelope (the left $OY$ axis). Dotted line indicates moment of the observation.}
    \label{fig:lightcurves}
\end{figure*}

\begin{figure}
\begin{center}
	\includegraphics[width=0.55\columnwidth]{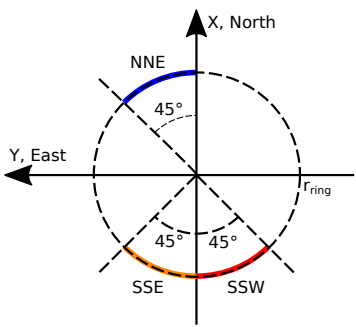}
\end{center}
    \caption{The model of V~CVn circumstellar envelope which was used for the approximation of DSP observables: three scattering arcs, each subtending angle of $45^{\circ}$, at fixed position angles. Elementary parts of the arcs are polarized perpendicularly to the direction to the star. 
    }
    \label{fig:trioctmodel}
\end{figure}

The data were reduced using the Differential Speckle Polarimetry (DSP) method described by \citet{SPPinstrument2}. As a result, we obtained estimations of the visibility ratios of the object in  orthogonally polarized light \citep{Norris2012}:
\begin{equation}
\mathcal{R}_{Q}(\boldsymbol{f}) = \frac{\widetilde{I}(\boldsymbol{f})+\widetilde{Q}(\boldsymbol{f})}{\widetilde{I}(\boldsymbol{f})-\widetilde{Q}(\boldsymbol{f})},\,\,\,\mathcal{R}_{U}(\boldsymbol{f}) = \frac{\widetilde{I}(\boldsymbol{f})+\widetilde{U}(\boldsymbol{f})}{\widetilde{I}(\boldsymbol{f})-\widetilde{U}(\boldsymbol{f})}.
\label{eq:RvalDef}
\end{equation}
Here $\widetilde{I}$, $\widetilde{Q}$, and $\widetilde{U}$ are the Fourier transforms of the Stokes parameters distributions in the object. $\boldsymbol{f}$ is the spatial frequency. As one can see, it is possible to define two ratios: $\mathcal{R}_Q$ and $\mathcal{R}_U$ for the Stokes parameters $Q$ and $U$, respectively. DSP allows to estimate both the amplitude and phase of $\mathcal{R}$ value. The observations were conducted at the Cassegrain and Nasmyth foci of the telescope. In the case of the Nasmyth focus the correction for the instrumental polarization effects was applied \citep{SPPinstrument2}. 

The measurements of $\mathcal{R}$ value for the two dates and two filters are presented in Fig.~\ref{fig:VCVn_RVal}. The polarized flux was clearly resolved for these cases as long as $\mathcal{R}$ values deviate from unity significantly.

\citet{SPPinstrument2} demonstrated how the distribution of Stokes parameters in some object can be estimated from the $\mathcal{R}$ value measurements. The polarized intensity computed using this method is displayed in the bottom row of Fig.~\ref{fig:VCVn_RVal}. 

The polarization of each elementary area of the envelope is roughly perpendicular to the direction connecting this area and the star. Therefore this envelope is likely to be a reflection nebula surrounding the star. The results of observations conducted on the same dates but at different filters are in good agreement. On the other hand, the difference between the observations conducted on two dates is striking. The nebula was dominated by the feature at north--northeast of the star on March 5th. 82 days later, on May 27th, the feature at south--southeast became significantly brighter than the northern one. For some other dates the feature at south--southwest became prominent. The images similar to Fig.~\ref{fig:VCVn_RVal} for all filters and all dates are provided at http://lnfm1.sai.msu.ru/kgo/mfc\_VCVn\_en.php.

The detected nebula has a characteristic angular size comparable to the diffraction--limited resolution of the telescope. The appearance of the images in polarized intensity is strongly affected by the blurring by the PSF. Due to this and some others problems discussed by \citet{SPPinstrument2}, these images can only be analyzed qualitatively. For the quantitative interpretation of the observations we will compare model and observations in terms of $\mathcal{R}$.


\section{MODELLING}
\label{sec:model}

Judging by the polarized intensity images of V~CVn for all dates, the simplest model for this source appears to be an unpolarized central star surrounded by three scattering arcs, see Fig.~\ref{fig:trioctmodel}. The configuration of the arcs is fixed. We will denote these arcs in accordance with their position relative to the star: north--northeast (NNE), south--southeast (SSE), and south-southwest (SSW). The radius is assumed to be the same for all arcs. Each elementary interval of the arcs is polarized perpendicularly to the direction connecting this interval and the star. 

In the frame of this model a single observation can be described by 4 parameters: the radius of the arcs $r_\mathrm{e}$ and ratios of their polarized fluxes to the total flux of the object: $F_{\mathrm{NNE}}, F_{\mathrm{SSE}}, F_{\mathrm{SSW}}$. The polarized flux is the product of polarization fraction and the total flux.  These parameters defines the distribution of Stokes parameters in the object, from which we computed the expected $\mathcal{R}_Q$ and $\mathcal{R}_U$ (\ref{eq:RvalDef}). We emphasize that in the frame of method which we use it is impossible to estimate the polarization fraction of arc and its total flux. We can estimate only their product: polarized flux.

To compare the modelled and observed $\mathcal{R}$ values,  we calculated a total residual weighted by $\sigma^{-2}$, where $\sigma$  is the uncertainty of $\mathcal{R}$ determination \citep{SPPinstrument2}. The residual was summed in the frequency domain, where the signal--to--noise ratio in $\mathcal{R}$ is sufficiently high, see Fig.~\ref{fig:VCVn_RVal}. The optimal model was found by minimizing the total residual.

We assumed that the arc's radius $r_\mathrm{e}$ does not depend on time and wavelength. It was determined by the approximation of the observations at three filters conducted on two dates: 2018 Mar 6th and 2018 May 27th. We found the radius to be $35\pm1$~mas. As long as $r_\mathrm{e}$ is less than formal diffration limited resolution of the telescope, its value is model--dependent. For example, it would be different for the model of sectors, not arcs. Nevertheless the departure of polarized envelope from the point--like star is detected quite reliably, and $r_\mathrm{e}$ can be considered as its characteristic angular extent.
 
We fixed the radius of arcs at $35$~mas and approximated each observation individually varying the rest three parameters $F_{\mathrm{NNE}}, F_{\mathrm{SSE}}, F_{\mathrm{SSW}}$. The examples of modelled $\mathcal{R}$ values and the corresponding images are displayed in Fig.~\ref{fig:VCVn_RVal}. The agreement with the observations is reasonable. The results of approximation of all observations are provided in Table \ref{table:model}. The $\chi_r^2$ statistics is less than 3 for 39 observations out of 57. Thus for most observations the model describes the observations satisfactorily. 





The total polarized flux from the envelope amounts to 0.01-0.03 of total flux of the object. In other words, the envelope explains the polarization of the object, while the unpolarized flux from the star dominates total the flux.

The Fig.~\ref{fig:lightcurves} displays the behaviour of the components of the envelope in $V$ and 550 filters. For ease of comparison, the AAVSO light curve \citep{Kafka2018} is given on the same graph. Note that the polarized fluxes of the envelope's arcs are normalized by the average flux of the object. This allows to inspect the variability of the arcs independently of the variability of the star.

It follows from Fig.~\ref{fig:lightcurves} that the brightening of the arc NNE coincides with the minimum brightness of the whole object. This leads to rise of the total polarization of the object. At the same time the arcs SSE and SSW change their brightness synchronously with the star. 

\begin{table*}
\begin{center}
\caption{
The results of modelling of speckle polarimetric observations of V~CVn: polarized fluxes of envelope components relative to the total flux of the object for the moment of respective observation. JD is given at series center. Magnitude was estimated by quasi--simultaneous observations of nearby non--variable star HIP65550. For HIP65550 we adopted the following magnitudes in bands 550, 625 and 880 nm: 5.9, 5.1 and 4.9, respectively. The degree and fraction of polarization was estimated from the series used for speckle polarimetric processing. The uncertainties are provided in 1$\sigma$ sense. $\chi_r^2$ characterizes goodness--of--fit. \label{table:model}}
\tabcolsep=0.12cm
\begin{tabular}{ccccccccc}
\hline
          &                  &             &        & & \multicolumn{3}{c}{polarized flux, $10^{-2}$}  & \\
JD        & filter           & mag         & $p$,\% & $\theta, ^{\circ}$ & NNE & SSE & SSW & $\chi_r^2$ \\
\hline
2457818.4 & $V$ & 8.00 & $2.77\pm0.15$ & $116\pm3$ & $1.85\substack{+0.12 \\ -0.32}$ & $0.08\substack{+0.19 \\ -0.07}$ & $0.74\substack{+0.08 \\ -0.17}$ & 2.1        \\
          & $I$ & 3.80 & $1.19\pm0.15$ & $110\pm7$ & $0.79\substack{+0.32 \\ -0.21}$ & $0.10\substack{+0.15 \\ -0.05}$ & $0.26\substack{+0.38 \\ -0.20}$ & 13.9       \\
2457886.2 & $V$ & 7.50 & $1.48\pm0.15$ & $107\pm6$ & $0.30\substack{+0.12 \\ -0.04}$ & $0.47\substack{+0.14 \\ -0.04}$ & $1.04\substack{+0.04 \\ -0.08}$ & 1.9        \\
          & $I$ & 3.60 & $0.67\pm0.15$ & $97\pm13$ & $0.38\substack{+0.26 \\ -0.18}$ & $0.25\substack{+0.19 \\ -0.03}$ & $0.60\substack{+0.20 \\ -0.15}$ & 9.6        \\
2458090.6 & $550$ & 7.10 & $0.90\pm0.15$ & $106\pm10$ & $0.22\substack{+0.02 \\ -0.02}$ & $0.35\substack{+0.02 \\ -0.02}$ & $0.66\substack{+0.02 \\ -0.02}$ & 1.7     \\
          & $625$ & 5.90 & $0.96\pm0.15$ & $107\pm9$ & $0.26\substack{+0.03 \\ -0.01}$ & $0.28\substack{+0.03 \\ -0.01}$ & $0.62\substack{+0.02 \\ -0.01}$ & 2.0      \\
          & $880$ & 3.70 & $0.70\pm0.15$ & $104\pm12$ & $0.27\substack{+0.09 \\ -0.01}$ & $0.14\substack{+0.05 \\ -0.01}$ & $0.38\substack{+0.04 \\ -0.01}$ & 3.0     \\
2458138.6 & $550$ & 7.10 & $1.32\pm0.15$ & $107\pm7$ & $0.86\substack{+0.05 \\ -0.05}$ & $0.24\substack{+0.05 \\ -0.04}$ & $0.36\substack{+0.05 \\ -0.05}$ & 9.0      \\
          & $625$ & 5.80 & $1.33\pm0.15$ & $102\pm6$ & $0.73\substack{+0.03 \\ -0.04}$ & $0.09\substack{+0.04 \\ -0.03}$ & $0.40\substack{+0.04 \\ -0.04}$ & 0.9      \\
          & $880$ & 3.30 & $1.05\pm0.15$ & $96\pm8$ & $0.44\substack{+0.06 \\ -0.01}$ & $0.16\substack{+0.04 \\ -0.01}$ & $0.38\substack{+0.04 \\ -0.01}$ & 2.8       \\
2458156.6 & $550$ & 7.80 & $1.15\pm0.15$ & $106\pm7$ & $0.63\substack{+0.03 \\ -0.03}$ & $0.18\substack{+0.03 \\ -0.03}$ & $0.45\substack{+0.03 \\ -0.03}$ & 1.6      \\
          & $625$ & 6.40 & $1.31\pm0.15$ & $104\pm7$ & $0.74\substack{+0.02 \\ -0.02}$ & $0.33\substack{+0.02 \\ -0.02}$ & $0.53\substack{+0.03 \\ -0.02}$ & 1.6      \\
          & $880$ & 3.70 & $0.96\pm0.15$ & $104\pm9$ & $0.52\substack{+0.04 \\ -0.02}$ & $0.22\substack{+0.02 \\ -0.01}$ & $0.34\substack{+0.03 \\ -0.02}$ & 6.3      \\
2458176.6 & $550$ & 7.50 & $1.46\pm0.15$ & $109\pm6$ & $1.01\substack{+0.03 \\ -0.04}$ & $0.04\substack{+0.05 \\ -0.03}$ & $0.46\substack{+0.03 \\ -0.03}$ & 1.3      \\
          & $625$ & 6.30 & $1.54\pm0.15$ & $105\pm6$ & $1.05\substack{+0.02 \\ -0.03}$ & $0.23\substack{+0.03 \\ -0.02}$ & $0.40\substack{+0.02 \\ -0.03}$ & 1.5      \\
          & $880$ & 3.70 & $1.10\pm0.15$ & $102\pm8$ & $0.65\substack{+0.01 \\ -0.01}$ & $0.21\substack{+0.01 \\ -0.01}$ & $0.30\substack{+0.02 \\ -0.01}$ & 4.2      \\
2458183.6 & $550$ & 7.40 & $2.03\pm0.15$ & $108\pm4$ & $1.58\substack{+0.05 \\ -0.06}$ & $0.10\substack{+0.06 \\ -0.04}$ & $0.57\substack{+0.05 \\ -0.06}$ & 1.0      \\
          & $625$ & 6.30 & $1.99\pm0.15$ & $106\pm4$ & $1.44\substack{+0.03 \\ -0.04}$ & $0.27\substack{+0.04 \\ -0.03}$ & $0.47\substack{+0.03 \\ -0.03}$ & 1.1      \\
          & $880$ & 4.00 & $1.16\pm0.15$ & $104\pm7$ & $0.72\substack{+0.02 \\ -0.02}$ & $0.15\substack{+0.03 \\ -0.01}$ & $0.33\substack{+0.03 \\ -0.02}$ & 1.9      \\
2458210.3 & $550$ & 7.70 & $2.80\pm0.15$ & $116\pm3$ & $1.93\substack{+0.05 \\ -0.12}$ & $0.01\substack{+0.06 \\ -0.01}$ & $0.89\substack{+0.05 \\ -0.07}$ & 1.5      \\
          & $625$ & 6.80 & $2.52\pm0.15$ & $113\pm3$ & $1.72\substack{+0.03 \\ -0.12}$ & $0.01\substack{+0.10 \\ -0.01}$ & $0.83\substack{+0.04 \\ -0.10}$ & 1.4      \\
          & $880$ & 4.00 & $0.86\pm0.15$ & $106\pm10$ & $0.61\substack{+0.02 \\ -0.02}$ & $0.08\substack{+0.04 \\ -0.01}$ & $0.20\substack{+0.04 \\ -0.01}$ & 2.9     \\
2458234.4 & $550$ & 7.50 & $0.64\pm0.15$ & $112\pm13$ & $0.21\substack{+0.05 \\ -0.04}$ & $0.43\substack{+0.03 \\ -0.04}$ & $0.28\substack{+0.05 \\ -0.04}$ & 2.8     \\
          & $625$ & 6.40 & $1.06\pm0.15$ & $107\pm8$ & $0.49\substack{+0.04 \\ -0.03}$ & $0.37\substack{+0.03 \\ -0.03}$ & $0.50\substack{+0.04 \\ -0.03}$ & 2.3      \\
          & $880$ & 3.60 & $0.81\pm0.15$ & $95\pm11$ & $0.33\substack{+0.03 \\ -0.02}$ & $0.38\substack{+0.02 \\ -0.01}$ & $0.32\substack{+0.02 \\ -0.01}$ & 5.8      \\
2458238.4 & $550$ & 7.40 & $0.39\pm0.15$ & $96\pm22$ & $0.12\substack{+0.07 \\ -0.03}$ & $0.46\substack{+0.03 \\ -0.03}$ & $0.07\substack{+0.06 \\ -0.02}$ & 3.9      \\
          & $625$ & 5.90 & $0.88\pm0.15$ & $102\pm10$ & $0.36\substack{+0.05 \\ -0.04}$ & $0.46\substack{+0.04 \\ -0.04}$ & $0.39\substack{+0.04 \\ -0.04}$ & 2.7     \\
          & $880$ & 3.50 & $0.80\pm0.15$ & $93\pm11$ & $0.27\substack{+0.03 \\ -0.02}$ & $0.40\substack{+0.02 \\ -0.02}$ & $0.37\substack{+0.04 \\ -0.02}$ & 8.2      \\
2458254.4 & $550$ & 6.80 & $0.95\pm0.15$ & $72\pm9$ & $0.01\substack{+0.04 \\ -0.01}$ & $0.90\substack{+0.04 \\ -0.04}$ & $0.29\substack{+0.07 \\ -0.04}$ & 1.2       \\
          & $625$ & 5.60 & $1.03\pm0.15$ & $78\pm8$ & $0.04\substack{+0.06 \\ -0.02}$ & $0.77\substack{+0.03 \\ -0.02}$ & $0.31\substack{+0.06 \\ -0.02}$ & 1.4       \\
          & $880$ & 3.30 & $0.93\pm0.15$ & $84\pm9$ & $0.14\substack{+0.02 \\ -0.01}$ & $0.58\substack{+0.02 \\ -0.01}$ & $0.39\substack{+0.01 \\ -0.01}$ & 2.9       \\
2458266.4 & $550$ & 6.90 & $1.31\pm0.15$ & $88\pm7$ & $0.15\substack{+0.04 \\ -0.02}$ & $0.79\substack{+0.03 \\ -0.02}$ & $0.69\substack{+0.02 \\ -0.02}$ & 2.6       \\
          & $625$ & 5.50 & $1.18\pm0.15$ & $85\pm7$ & $0.11\substack{+0.06 \\ -0.01}$ & $0.69\substack{+0.01 \\ -0.01}$ & $0.53\substack{+0.03 \\ -0.01}$ & 3.3       \\
          & $880$ & 3.60 & $0.86\pm0.15$ & $95\pm10$ & $0.26\substack{+0.04 \\ -0.01}$ & $0.29\substack{+0.03 \\ -0.01}$ & $0.45\substack{+0.02 \\ -0.02}$ & 3.2      \\
2458275.3 & $550$ & 6.80 & $1.21\pm0.15$ & $92\pm7$ & $0.30\substack{+0.02 \\ -0.02}$ & $0.65\substack{+0.02 \\ -0.02}$ & $0.59\substack{+0.02 \\ -0.02}$ & 1.4       \\
          & $625$ & 5.70 & $1.38\pm0.15$ & $94\pm6$ & $0.42\substack{+0.02 \\ -0.02}$ & $0.46\substack{+0.02 \\ -0.02}$ & $0.75\substack{+0.02 \\ -0.02}$ & 1.6       \\
          & $880$ & 3.60 & $1.05\pm0.15$ & $94\pm8$ & $0.35\substack{+0.02 \\ -0.01}$ & $0.41\substack{+0.02 \\ -0.01}$ & $0.54\substack{+0.01 \\ -0.01}$ & 4.8       \\
2458326.2 & $550$ & 7.00 & $1.43\pm0.15$ & $97\pm6$ & $0.48\substack{+0.04 \\ -0.03}$ & $0.47\substack{+0.04 \\ -0.03}$ & $0.64\substack{+0.04 \\ -0.03}$ & 1.2       \\
          & $625$ & 6.10 & $1.59\pm0.15$ & $93\pm5$ & $0.58\substack{+0.02 \\ -0.02}$ & $0.65\substack{+0.02 \\ -0.02}$ & $0.63\substack{+0.03 \\ -0.02}$ & 2.4       \\
          & $880$ & 3.60 & $1.30\pm0.15$ & $87\pm7$ & $0.46\substack{+0.02 \\ -0.02}$ & $0.29\substack{+0.02 \\ -0.01}$ & $0.38\substack{+0.02 \\ -0.01}$ & 15.8      \\
2458336.2 & $550$ & 7.10 & $1.12\pm0.15$ & $100\pm8$ & $0.34\substack{+0.06 \\ -0.05}$ & $0.36\substack{+0.05 \\ -0.04}$ & $0.54\substack{+0.05 \\ -0.05}$ & 0.9      \\
          & $625$ & 6.00 & $1.29\pm0.15$ & $93\pm7$ & $0.46\substack{+0.03 \\ -0.02}$ & $0.48\substack{+0.02 \\ -0.02}$ & $0.48\substack{+0.03 \\ -0.02}$ & 1.2       \\
          & $880$ & 3.70 & $1.18\pm0.15$ & $85\pm7$ & $0.31\substack{+0.04 \\ -0.01}$ & $0.31\substack{+0.02 \\ -0.01}$ & $0.24\substack{+0.03 \\ -0.01}$ & 8.2       \\
2458356.2 & $550$ & 7.20 & $1.11\pm0.15$ & $102\pm8$ & $0.59\substack{+0.03 \\ -0.03}$ & $0.32\substack{+0.03 \\ -0.02}$ & $0.35\substack{+0.03 \\ -0.02}$ & 1.7      \\
          & $625$ & 6.00 & $1.26\pm0.15$ & $94\pm7$ & $0.55\substack{+0.02 \\ -0.02}$ & $0.59\substack{+0.02 \\ -0.02}$ & $0.38\substack{+0.03 \\ -0.02}$ & 2.5       \\
          & $880$ & 3.80 & $1.05\pm0.15$ & $84\pm8$ & $0.29\substack{+0.04 \\ -0.01}$ & $0.33\substack{+0.02 \\ -0.01}$ & $0.18\substack{+0.04 \\ -0.01}$ & 11.4      \\
2458375.2 & $550$ & 7.40 & $1.56\pm0.15$ & $104\pm6$ & $1.01\substack{+0.04 \\ -0.04}$ & $0.22\substack{+0.04 \\ -0.04}$ & $0.40\substack{+0.05 \\ -0.04}$ & 1.6      \\
          & $625$ & 6.20 & $1.41\pm0.15$ & $97\pm6$ & $0.84\substack{+0.03 \\ -0.02}$ & $0.38\substack{+0.03 \\ -0.03}$ & $0.26\substack{+0.04 \\ -0.02}$ & 1.9       \\
          & $880$ & 3.90 & $0.89\pm0.15$ & $82\pm10$ & $0.30\substack{+0.05 \\ -0.02}$ & $0.25\substack{+0.03 \\ -0.02}$ & $0.04\substack{+0.06 \\ -0.02}$ & 5.2      \\
2458446.5 & $550$ & 7.00 & $1.01\pm0.15$ & $86\pm9$ & $0.08\substack{+0.23 \\ -0.06}$ & $0.57\substack{+0.11 \\ -0.05}$ & $0.58\substack{+0.08 \\ -0.06}$ & 1.5       \\
          & $625$ & 6.00 & $1.11\pm0.15$ & $98\pm8$ & $0.30\substack{+0.08 \\ -0.03}$ & $0.46\substack{+0.08 \\ -0.03}$ & $0.67\substack{+0.03 \\ -0.09}$ & 1.4       \\
          & $880$ & 3.60 & $1.30\pm0.15$ & $103\pm7$ & $0.39\substack{+0.14 \\ -0.04}$ & $0.35\substack{+0.10 \\ -0.03}$ & $0.78\substack{+0.04 \\ -0.10}$ & 3.6      \\
2458459.6 & $550$ & 7.30 & $1.03\pm0.15$ & $95\pm8$ & $0.15\substack{+0.18 \\ -0.05}$ & $0.55\substack{+0.10 \\ -0.04}$ & $0.60\substack{+0.09 \\ -0.04}$ & 1.1       \\
          & $625$ & 6.00 & $1.38\pm0.15$ & $98\pm6$ & $0.50\substack{+0.09 \\ -0.04}$ & $0.56\substack{+0.06 \\ -0.03}$ & $0.75\substack{+0.03 \\ -0.07}$ & 1.4       \\
          & $880$ & 3.50 & $1.23\pm0.15$ & $98\pm7$ & $0.40\substack{+0.09 \\ -0.03}$ & $0.45\substack{+0.06 \\ -0.02}$ & $0.74\substack{+0.04 \\ -0.03}$ & 4.0       \\
2458504.5 & $550$ & 7.10 & $0.72\pm0.15$ & $95\pm12$ & $0.27\substack{+0.22 \\ -0.04}$ & $0.16\substack{+0.14 \\ -0.04}$ & $0.33\substack{+0.15 \\ -0.04}$ & 3.0      \\
          & $625$ & 6.00 & $0.81\pm0.15$ & $97\pm11$ & $0.30\substack{+0.12 \\ -0.04}$ & $0.15\substack{+0.19 \\ -0.04}$ & $0.41\substack{+0.12 \\ -0.04}$ & 3.2      \\
          & $880$ & 3.50 & $0.68\pm0.15$ & $95\pm13$ & $0.26\substack{+0.17 \\ -0.05}$ & $0.13\substack{+0.16 \\ -0.04}$ & $0.31\substack{+0.17 \\ -0.05}$ & 9.5      \\
\hline
\end{tabular}
\end{center}
\end{table*}

These features were roughly reproduced at the considered cycles of pulsation. The exact repetition is not expected anyway, because the pulsation of the star is irregular. For example, in the period of pulsation between JD=2458260 and JD=2458440 there was no prominent minimum of brightness on the light curve. The polarization fraction stayed below 2 per cent. The difference in arcs brightnesses in that period was less pronounced with respect to the previous minimum. Nevertheless the overall character of the arcs brightnesses variability was the same. 

\section{DISCUSSION}
\label{sec:interp}

In accordance with the latest measurements of \citet{Gaia2018} the distance to V~CVn is $1.27\pm0.24$~kpc. The star resides quite high above the galactic plane: $1.17\pm0.23$~kpc. The apparent proper motion of V~CVn is $\mu_\alpha\cos\delta=-38.99\pm0.20$~mas$\,$yr$^{-1}$ and $\mu_\delta=-11.768$~mas$\,$yr$^{-1}$, the correspoding tangential velocity is 251 km s$^{-1}$ with respect to the Sun. The radial velocity is 4.7 km$\,$s$^{-1}$ \citep{Famaey2009}, i.e. the star travels almost perpendicularly to the line of sight.

The parallax of V~CVn is determined with error of 0.14~mas, what is three times larger than the median of this value for stars with $G\approx8.5$ \citep{Lindegren2018}. The error of proper motion is relatively large as well. The excessive noise of astrometrical solution can be caused either by chromatic instrumental effects inherent to {\it Gaia} data, or by the envelope described in previous section.


\citet*{Sharma2016} provide the following fundamental parameters of the star: the spectral type is M6III, the effective temperature is $3180\pm99$~K. The luminosity corresponding to the {\it Gaia} distance is $3.6\pm1.5\times10^4\,L_\odot$ \citep*{McDonald2012}. One can also estimate the radius of the star: $R_\star=590\pm110\,R_\odot$, which results in the angular radius of 2.2~mas at the distance of the object.




The characteristic linear radius of the found nebula is $44\pm10$ au, what is $\approx10$ times larger than $R_\star$. Therefore  the observed polarized flux is formed in the circumstellar envelope at a significant distance from the photosphere. Most likely this envelope is generated by the dusty stellar wind. Mid--IR excess in the spectrum of V~CVn \citep{Price2010} and the 9.7~$\mu$m silicate feature also favour the existence of a dusty envelope \citep{Olnon1986,Simpson1991}. At the same time the star has quite low colour excess $E(B-V)=0.04$ \citep{Montez2017} which indicates that the envelope is not spherically symmetric.

The Keplerian motion is the most obvious potential explanation for the observed changes in the morphology of the envelope. However getting across the semicircle with the radius of $44$~au in half the period of pulsation (P$\approx$194~days \citealt{Samus2017}) requires a velocity of $\approx2500$~km$\,$s$^{-1}$. This is much larger than the expected Keplerian velocity at $44$~au from the star with the mass less than $10 M_\odot$. Therefore the hypothesis of Keplerian is rejected. In below we consider several other hypotheses for the interpretation of the V~CVn envelope.

\subsection{Bow shock hypothesis}
\label{subs:pm}


\citep{Neilson2014} proposed that the asymmetric dusty envelope could form behind the bowshock emerging at the boundary between stellar wind and the interstellar medium (ISM). The distance between the star and an apex of the bow shock is defined by the equality of ram pressures of the stellar wind and of the flow of the ISM gas \citep{vanBuren1988}. Now we estimate at which density of ISM the bow shock will be located at 44 au from the star.

First we need a mass--loss rate of the stellar wind. This value can be estimated from the pulsation period $P$ using the relation by \citet{DeBeck2010}. In the case of V~CVn the expected mass--loss rate is $2\times10^{-7}M_\odot\,$yr$^{-1}$. 

For the estimation of velocity of the star relative to the ISM we performed the correction for the differential rotation of the Galaxy and for the motion of the Sun towards the apex. We used the maser rotation curve, which is closest to the kinematics of the gas \citep{Rastorguev2017}. The velocity of the star with respect to the local rest frame is $V_\star\approx237$~km s$^{-1}$. The position angle of velocity vector projection on the image plane is $255^{\circ}$, this direction is indicated by the arrow in the lower left panel of the Fig.~\ref{fig:VCVn_RVal}.

Knowing the velocity of the star relative to ISM, and using equation (1) from \citep{vanBuren1988}, it is possible to derive the following dependence of the required ISM density $n_\mathrm{H}$ on the velocity of the stellar wind $V_\mathrm{w}$:
\begin{equation}
n_\mathrm{H} = 490 V_\mathrm{w}.
\end{equation}
Here $V_\mathrm{w}$ is expressed in km$\,$s$^{-1}$, and $n_\mathrm{H}$ is expressed in cm$^{-3}$. The density of ISM is expected to be $\approx 10^3$ cm$^{-3}$ at the terminal stellar wind velocity of a few km$\,$s$^{-1}$, which is typical for this type of stars.

At the same time the average density of ISM at 1.2~kpc above the galactic plane is $\approx3.3\times10^{-3}$~cm$^{-3}$ (``best estimate'' from fig.~10 in \citealt{Dickey1990}). In accordance with the map of \ion{H}{I} obtained by \citet{HI4PI} no significant molecular cloud exists towards V~CVn. We conclude that the ISM in the vicinity of V~CVn is by $\approx5$ orders of magnitude less dense than required to form the bow shock at 44 au from the star. 

In other words, the bowshock has to form at the distances much larger than the size of the detected envelope. But intrinsically spherical stellar wind should retain this symmetry up to distances where the bowshock is formed. Therefore in our case the interaction between stellar wind and ISM should not induce asymmetry of dusty envelope, and, moreover, any variability of its surface brightness. It is more likely that the shape of the envelope is caused by the anisotropy of the stellar mass loss.





\subsection{Light echo hypothesis}
\label{subs:lightecho}

The fact that the brightness of the NNE arc reaches its maximum value after $\Delta t \sim 95$ days after the  maximum of brightness of the star can be explained by the effect of the light echo, as in case of RS~Pup \citep{Kervella2008}. In this case the NNE arc should be at least $c\Delta t/2\sim8000$~au farther from us than the star. The characteristic linear size of the cloud can be estimated as NNE arc length: $\approx40$ au. The respective circumstellar cloud would intercept no more than $\approx1.4\times10^{-6}$ of stellar radiation. Some of this radiation would be absorbed and the rest would be scattered mostly forward. For the observer the cloud would be seen in back--scattering regime as a source of polarized flux at least $\sim10^{-6}$ fainter than the star, i.e. $10^2-10^3$ fainter than observed --- see Fig. \ref{fig:lightcurves}. It means that the brightness of the detected nebula is inconsistent with the light echo hypothesis.

\begin{figure}
\begin{center}
	\includegraphics[width=1.05\columnwidth]{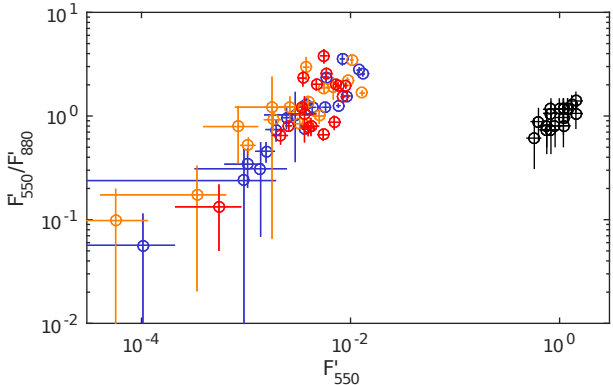}
\end{center}
    \caption{Colour--flux diagrams for envelope components and the whole object. Along $OX$ axis the flux in 550 filter normalized by the average total flux of the object is plotted. Along $OY$ axis the ratio of normalized fluxes in 550 and 880 filters is plotted.  Blue, orange and red symbols correspond to arc NNE, SSE and SSW, respectively. Black symbol corresponds to the whole object. Taking into account that the envelope is fainter than the star they represent the behaviour of the star. For the envelope components polarized fluxes are given, for the whole object total flux is given.
    }
    \label{fig:colorflux}
\end{figure}

\subsection{Variable shadowing hypothesis}

Periodic changes in morphology of circumstellar envelope may be caused by the effect of variable shadowing. In this case the outer parts of the circumstellar envelope will be partly obscured from the point of view of the star by the inner parts of the same nebula. The geometry of this obscuration will depend on the radius of the star, which in turn changes with pulsation. 

This hypothesis had been proposed by \citet{Kervella2014} to explain correlation between photocentre motion and pulsation cycle of another semiregular variable L$_2$~Pup. 

If this hypothesis is applicable to V~CVn, then there is a structure in the inner envelope of the star which casts a shadow on NNE arc when the star is bright. Assuming that the star is larger at minimum brighness, and shadowing of NNE arc should decrease when the object is faint. Consequently, NNE arc should become brigther.

However the colour (temperature) of star depends on the pulsation cycle as well. Therefore in the frame of this explanation the tracks of the arcs in the colour--flux diagram should differ. The NNE arc should become more red at its brightening (when the star is faint), while SSE/SSW should become more blue, when they brighten.

But it follows from Fig. \ref{fig:colorflux} that the colour behaviour of all components of the envelope is essentially the same and similar to one of the star: the bright state is characterized by the bluer colour. Therefore we reject this hypothesis as well.

\subsection{Non--radial pulsation hypothesis}

\begin{figure}
\includegraphics[width=\columnwidth]{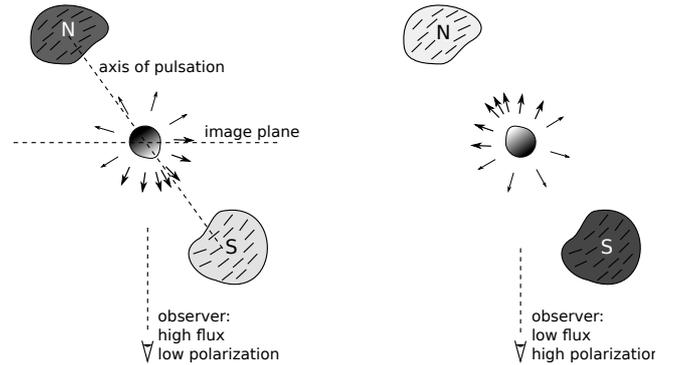}
\caption{The scheme illustrates a proposed configuration of the object in case of non--radial pulsation (not to scale). Left and right panels are for the maximum and minimum flux for the Earth--based observer, respectively. See the text for the explanation.}
\label{fig:scheme}
\end{figure}

The character of the variability of the NNE arc could be naturally explained in terms of the assumption that the pulsation of the star from its point of view is shifted by half a period relative to the pulsation from the point of view of the observer. At the same time from the point of view of SSW/SSE arcs the pulsation appears the same as for observer. The corresponding model is illustrated in Fig.\ref{fig:scheme} and assumes significant departure of stellar pulsation from purely radial, or, more precisely, the existence of dipolar pulsation.

In the frame of this model, when the star at maximum brigthness, the part of star facing the observer and arcs SSW/SSE is in bright state. Meanwhile the part of star facing the arc NNE is faint. After the half a period of pulsation the situation is opposite. Now the star appears faint for the observer and bright for the NNE arc. Because of this the latter reaches maximum brighness. The input of scattered and polarized radiation in the total flux of the object rises. The inverse correlation between the total fraction of polarization and flux emerges.

Non--radial pulsation was considered as one of possible qualitative explanations for the unusual polarization variability of the star V1497 Aql by \citet{Patel2008}. This semiregular variable star shows irregular changes in fraction of polarization with the amplitude of up to 5 per cent associated with relatively small changes in brighness $\approx0.2^m$. However, reliable evidences for non-radial pulsation giving an amplitude of $\Delta V \approx 1.5^m$ are missing and theoretically it was not predicted \citep{Mosser2013}.

\section{Conclusions}
\label{sec:conclusions}

We have presented the observations of the semiregular variable V~CVn using the method of differential speckle polarimetry at the wavelengths  of 550, 625 and 880~nm. We found a reflection nebula in polarized light surrounding the star at the typical distance of 35~mas, which corresponds to 44 au at the distance of the object. The detected nebula lacks the rotational symmetry in the plane of sky. Three regions can be identified in this nebula, towards north--northeast, south--southeast, and south-southwest from the star. The asymmetry of the nebula leads to the constancy of angle of polarization and to the high fraction of polarization for the whole object.

The observations on 20 dates distributed over the three cycles of the pulsation demonstrated that the different regions of the scattering envelope change their brightness with the same period as the star, but with significant phase shifts. For example, the region NNE reaches the maximum brightness when the whole object is at minimum brightness. At that time, the input of scattered, and therefore polarized, radiation in the total flux of the object increases. Because of this the whole object demonstrates an inverse correlation between the flux and polarization.

Using a simple estimation we have shown that the asymmetry of the envelope cannot be generated by the interaction of the stellar wind and ISM. The geometry of the envelope is likely to be defined solely by the anisotropy of the stellar wind. 

We demonstrate that the very peculiar variations of surface brightness of the envelope cannot be explained by Keplerian motion, light echo or variable shadowing. We note that all observational features of dusty envelope of V~CVn are in agreement with the assumption that the pulsation of the star is significantly non--radial. We leave the question whether such explanation is realistic from the point of view of stellar models open.

New observations of the envelope at angular resolution smaller than its characteristic size are needed for more detailed modelling of the envelope. Such observations could be conducted at a large telescope or a long--baseline interferometer. Both single and multi--epoch observations would be of value. Spectroscopic monitoring would allow to check the atmosphere for the temperature inhomogeneity across the stellar surface. 











\section*{Acknowledgements}

We are grateful to the staff of the Caucasian Observatory of SAI MSU for the help with conducting of observations used in this study. We thank the referee for the valuable comments. We acknowledge the variable star observations from the AAVSO International Database contributed by observers worldwide and used in this research. We acknowledge financial support from the Russian Foundation for Basic Research, project no. 16-32-60065 (BS --- observations and data processing) and no. 19-02-00611 (AR --- interpretation). AD acknowledges the support from the Program of development of M.V. Lomonosov Moscow State University (Leading Scientific School ``Physics of stars, relativistic objects and galaxies'').



\bibliographystyle{mnras}
\bibliography{VCVn} 





\bsp	
\label{lastpage}
\end{document}